\documentclass{kapproc/kapproc} 

\usepackage{kapproc/procps} 

\usepackage[dvips]{graphicx}

\upperandlowercase

\kluwerbib 

\begin{document}

\newcommand{\Sauron}{{\tt SAURON}}
\newcommand{\XSauron}{{\tt XS}auron}
\newcommand{\tiger}{{\tt TIGER}}
\newcommand{\rotcur}{{\tt ROTCUR}}
\newcommand{\gipsy}{{\tt GIPSY}}
\newcommand{\SN}{$S/N$}
\newcommand{\AN}{$A/N$}
\newcommand{\Ha}{H$\alpha$}
\newcommand{\Hb}{H$\beta$}
\newcommand{\Hg}{H$\gamma$}
\newcommand{\lda}{$\lambda$}
\newcommand{\mgb}{Mg$b$}
\newcommand{\mgt}{Mg$_2$}
\newcommand{\OI}{[{\sc O$\,$i}]}
\newcommand{\OII}{[{\sc O$\,$ii}]}
\newcommand{\OIII}{[{\sc O$\,$iii}]}
\newcommand{\NI}{[{\sc N$\,$i}]}
\newcommand{\NII}{[{\sc N$\,$ii}]}
\newcommand{\NeIII}{[{\sc Ne$\,$iii}]}
\newcommand{\SII}{[{\sc S$\,$ii}]}
\newcommand{\HI}{{\sc H$\,$i}}
\newcommand{\HII}{{\sc H$\,$ii}}
\newcommand{\plm}{$\, \pm \,$}
\newcommand{\Vsys}{$V_\mathrm{sys}$}
\newcommand{\Vrot}{$V_\mathrm{rot}$}
\newcommand{\Vrad}{$V_\mathrm{rad}$}
\newcommand{\epot}{$\varepsilon$}

\newcommand{\czero}{$c_0$}
\newcommand{\sone}{$s_1$}
\newcommand{\cone}{$c_1$}
\newcommand{\stwo}{$s_2$}
\newcommand{\ctwo}{$c_2$}
\newcommand{\sthree}{$s_3$}
\newcommand{\cthree}{$c_3$}

\def\kms{$\mbox{km s}^{-1}$}
\def\Myr{$\mbox{M}_\odot\mbox{ yr}^{-1}$}
\def\etal{et al.~}
\def\deg{^\circ}
\def\asim{\mathord{\sim}}
\def\farcs{\hbox{$.\!\!^{\prime\prime}$}}
\newcommand\arcmin{\hbox{$^\prime$}}
\newcommand\arcsec{\hbox{$^{\prime\prime}$}}
\newcommand\farcm{\hbox{$.\mkern-4mu^\prime$}}

\def\JF#1{{\bf {\sc JFB Comment:} #1}}
\newcommand{\refer}{({\sc References}) }
\newcommand{\KF}{{\sc KF Comment: }}
\newcommand{\EE}[1]{{\sc EE Comment: #1 }}



\articletitle[Two-dimensional Kinematics of a Bar and Central Disk in NGC~5448]
{Two-dimensional Kinematics of a Bar and Central Disk in NGC~5448}

\chaptitlerunninghead{Two-dimensional Kinematics of a Bar and Central Disk in NGC~5448}












\author{	Kambiz Fathi\altaffilmark{1,2},
		Glenn van de Ven\altaffilmark{3}, 
		Reynier Peletier\altaffilmark{2},
		Eric Emsellem\altaffilmark{4},
		Jes\'us Falc\'on--Barroso\altaffilmark{3},
		Michele Cappellari\altaffilmark{3}, 
		Tim de Zeeuw\altaffilmark{3}
		}

\altaffiltext{1}{Rochester Institute of Technology, 85 Lomb Memorial Drive, Rochester, New York 14623, USA}
\altaffiltext{2}{Kapteyn Astronomical Institute, Postbus 800, 9700 AV, Groningen, The Netherlands}
\altaffiltext{3}{Sterrewacht Leiden, Niels Bohrweg~2, 2333~CA Leiden, The Netherlands}
\altaffiltext{4}{Centre de Recherche Astronomique de Lyon, F-69561 Saint Genis Laval Cedex, France} 

%


\begin{abstract}
We analyse \Sauron\ kinematic maps of the inner kpc of the 
early-type (Sa) barred spiral galaxy NGC~5448. 
The observed morphology and kinematics of the emission-line gas
is patchy and perturbed, indicating clear departures from circular 
motion. The kinematics of the stars is more regular, and displays 
a small inner disk-like system embedded in a large-scale rotating 
structure. We focus on the \OIII\ gas, and use a harmonic 
decomposition formalism to analyse the gas velocity field. 
The higher-order harmonic terms and the main kinematic features 
of the observed data are consistent with an analytically 
constructed simple bar model, which is derived using linear theory. 
Our study illustrates how the harmonic decomposition formalism 
can be used as a powerful tool to quantify non-circular motions 
in observed gas velocity fields.
\end{abstract}


\section{Introduction}
\label{sec:5448intro}
Line-of-sight velocity distributions are efficient probes of the
dynamical structure of galaxies, and can be used to derive the mass
distribution, intrinsic shape and the motions of stars and gas.
Non-axisymmetric components such as bars, or external triggers, can
lead to significant galaxy evolution via, e.g., redistribution of the
angular momentum, triggering of star formation, or building of a
central mass concentration (e.g., Shloshman \etal 1989; Knapen \etal 2000). 
Much can be learned about these processes by
simultaneously studying both the stellar and gas dynamics of nearby
galaxies by means of integral-field spectroscopy. Here we summarize
such a study for the spiral galaxy NGC~5448, which was observed with
\Sauron\ as part of a representative survey of 72 nearby
early-type galaxies (de Zeeuw \etal 2002; Falc\'on--Barroso et al.\
2005). A more comprehensive report can be found in Fathi \etal (2005).


\section{Data}
\label{sec:sauron}
\begin{figure}
\center \includegraphics[scale=1.15,trim=0.0cm 0.0cm 0.0cm 0.0cm]{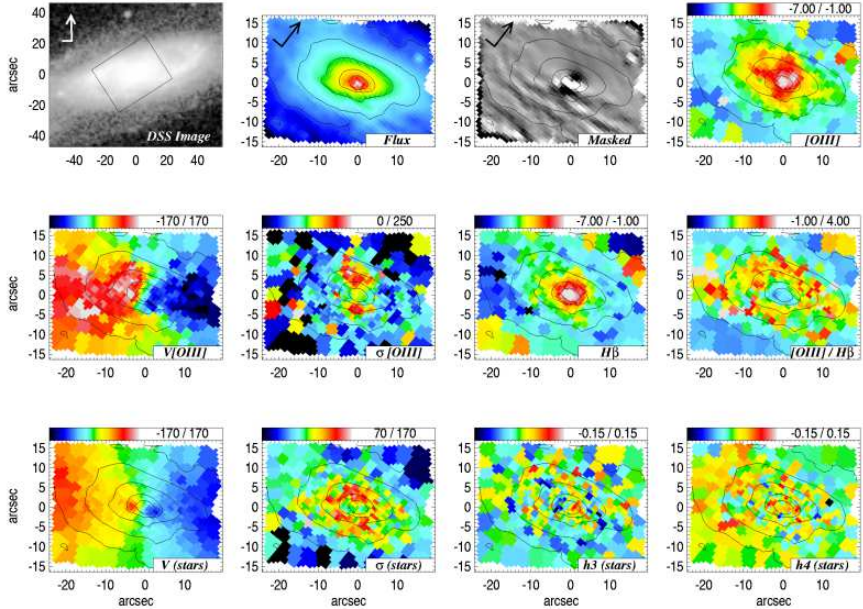} 
\caption{Top Left: Digitised Sky Survey image with \Sauron\ footprint and
north-east orientation arrow. All other panels show the \Sauron\ data with 
the same orientation and overplotted contours in magnitude steps of 0.25. 
Two-dimensional Voronoi binning has been applied to all our maps, as described
in Falc\'on--Barroso \etal (2005). The titles are indicated at the bottom right
corner of each panel with plotting ranges according to the top color bar. 
All velocities and velocity dispersions are given in \kms.}
\label{fig:datamaps}
\end{figure}

We observed NGC~5448 with \Sauron\ on April 14th 2004. 
Detailed specifications for the instrument, 
reduction procedure, and the data preparation procedure can be found 
in Bacon \etal (2001); Emsellem \etal (2004); Sarzi \etal (2005). 
The \Sauron\ flux map in Fig.~\ref{fig:datamaps} displays a smooth stellar 
distribution and the presence of prominent dust lanes to the south of the nucleus. 
The stellar kinematics shows a global disk-like rotation with 
a smaller inner stellar disk within the central 7\arcsec\ radius. 
Along the strong dust lanes, the gas shows a patchy distribution, 
with an asymmetric elongation of \OIII\ gas towards 
the east as well as the galactic poles. 
The gas velocity map clearly shows a prominent `S'-shaped zero-velocity 
curve with sharp edges indicating very strong non-circular gas motions.

\section{Analysis and Results}
\label{sec:sauron}
\begin{figure}
\center \includegraphics[scale=0.75,trim=0.0cm 0.0cm 0.0cm 0.0cm]{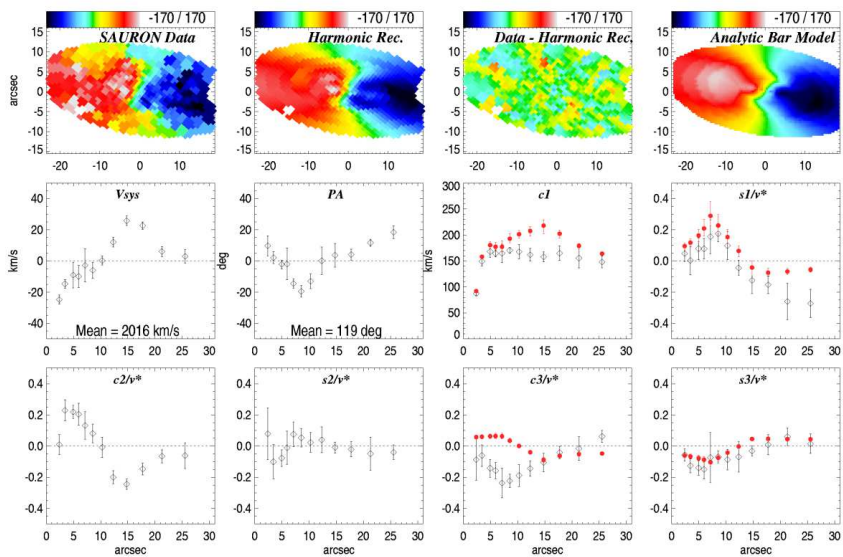} 
\caption{Top row, from left to right: observed \Sauron\
gas velocity field of NGC~5448; the
harmonic reconstruction of the \Sauron\ gas velocity field;
residual field; and the analytic
bar model, which reproduces best the main kinematic features of
the observed gas velocity field (in \kms). Middle and bottom row: the
tilted-ring and harmonic parameters as a function of outer radius of each ring,
and $v^* =c_1 \sin i$. The over-plotted red filled circles are the
analytically calculated first and third harmonic terms for the bar
model, with the $3\sigma$ (99.7\%) confidence level error bars.
The orientation of the maps is the same as in Fig. \ref{fig:datamaps}.}
\label{fig:analysis}
\end{figure}

In order to study the gas velocity field, we apply the
tilted-ring decomposition combined with the harmonic 
expansion formalism from Schoenmakers \etal (1997). 
This formalism allows us to extract the gaseous 
kinematic information such as the rotation 
curve, kinematic position angle variation, and higher harmonic terms. 
Wong \etal (2004) derived the higher-order harmonics for an axisymmetric 
potential with an $m=2$ bar perturbation. This bar model depends on 
a set of parameters describing the potential, as well as the viewing 
angle. We construct libraries of models with different input parameters, 
and find that we can describe our observed velocity field with 
the bar model shown in Fig.~\ref{fig:analysis}. Thus, the radial 
motions of the gas are associated with that of the large-scale bar 
(see Fathi \etal 2005 for a detailed discussion).

\begin{figure}
\center \includegraphics[scale=1.15,trim=0.5cm 0.0cm 0.0cm 0.0cm, angle=0]{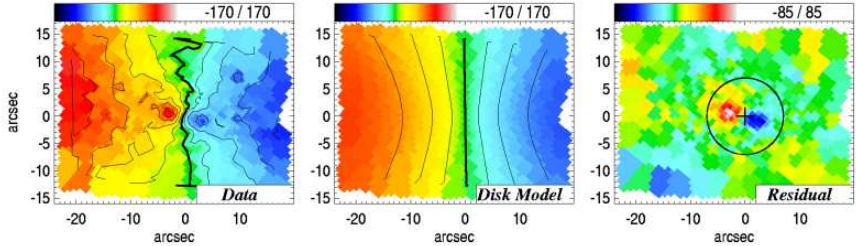} 
\caption{A thin isothermal disk model for the stellar velocity
field of NGC~5448. The circle marks the 7\arcsec\ region within
which we find a disk-like structure. All maps are given in \kms.}
\label{fig:diskmodel}
\end{figure}

NGC~5448 exhibits clear signatures of the presence of other components 
than a single bar, which affect the observed velocity field. 
Inspecting the photometry and the central parts of the gaseous 
and stellar velocity field, we find signatures of a central  
stellar disk embedded in the larger disk (see Fig.~\ref{fig:diskmodel}). 
We approximate the stellar velocity field with an exponential 
thin disk to emphasize the kinematic signatures of 
the central disk. Fig.~\ref{fig:diskmodel} illustrates our simple 
inclined disk model, after subtraction of which we find that the 
inner stellar disk-like component rotates faster than the outer disk. 
Fitting also an exponential disk to the region interior to the 7\arcsec\ 
radius, we find that the central disk-like structure is misaligned 
with respect to the outer disk by $\sim 13\deg$. 
The stellar kinematic maps show that the central disk rotates faster 
than the main disk, and our observed gas distribution and kinematics 
indicate that this central disk also hosts gas which rotates 
faster than its stellar counterpart. It is known that bars are 
efficient in transferring mass towards the inner regions of their 
host galaxies. The centrally concentrated matter may be able to form 
a secondary bar or a central disk. 
Our analysis shows that the non-circular gas kinematics 
in NGC~5448 could be driven by the large-scale bar. The central disk could
then have been formed as a result of the gas accumulation at the centre.\\

It is a pleasure to acknowledge the entire \Sauron\ team for their 
efforts in carrying out the observations and data preparation and for 
many fruitful discussions. KF thanks the conference organizers for all their efforts.



%


\bibliographystyle{kapalike}
\chapbblname{}
\chapbibliography{}


\end{document}